# Design and Development of Laughter Recognition System Based on Multimodal Fusion and Deep Learning

**Fuzheng Zhao [1*], Yu Bai [23]**

[1]Jilin University, China // [2]Kyushu University, Japan //// [3]Northeast Normal University, China
zhaofz635@gmail.com // Baiy173@nenu.edu.cn

**ABSTRACT:** This study aims to design and implement a laughter recognition system based on multimodal fusion and deep learning, leveraging image and audio processing technologies to achieve accurate laughter recognition and emotion analysis. First, the system loads video files and uses the OpenCV library to extract facial information while employing the Librosa library to process audio features such as MFCC. Then, multimodal fusion techniques are used to integrate image and audio features, followed by training and prediction using deep learning models. Evaluation results indicate that the model achieved 80% accuracy, precision, and recall on the test dataset, with an F1 score of 80%, demonstrating robust performance and the ability to handle real-world data variability. This study not only verifies the effectiveness of multimodal fusion methods in laughter recognition but also highlights their potential applications in affective computing and human-computer interaction. Future work will focus on further optimizing feature extraction and model architecture to improve recognition accuracy and expand application scenarios, promoting the development of laughter recognition technology in fields such as mental health monitoring and educational activity evaluation.

## 1. Introduction

Laughter, as a vital component of human emotional expression, is not only prevalent in daily life but also plays a significant role in facilitating emotional communication, enhancing social connections, reducing stress, and improving mental health. Provine (2001) delves into the scientific study of laughter, thoroughly exploring its origins, functions, and impacts, highlighting its importance in social interactions and its positive effects on physical and mental well-being. Martin and Ford (2018) comprehensively synthesizes research findings in psychology related to humor and laughter, further elucidating the psychological mechanisms, social functions, and individual mental health benefits of laughter (Martin, 2007). Kuiper and Martin (1988) indicate that laughter in daily life is closely tied to positive emotions, effectively reducing negative emotions and coping with stress, thereby exerting a positive impact on mental health.

In educational settings, students' emotional experiences and enjoyment cannot be overlooked. By identifying and utilizing these emotional elements, teaching effectiveness and students' learning experiences can be significantly enhanced. For instance, organizing diverse student activities based on their emotional experiences, such as symposia, team projects, and cultural events, not only promotes interaction and communication among students but also strengthens class cohesion and friendship (Goddard, 2003). Furthermore, selecting teachers with a strong sense of humor based on students' emotional feedback fosters a more positive and vibrant learning environment, boosting students' motivation and engagement (Martin et al., 2003). Encouraging students to vote for the funniest stories and jokes also shares joy and humor, enhancing mutual understanding and trust among them (Kuiper & Martin, 1998).

However, current laughter recognition methods face numerous challenges. Firstly, laughter datasets are scarcer and more heterogeneous compared to other speech datasets, significantly limiting the training and evaluation of laughter recognition models (Turker et al., 2017). Secondly, laughter exhibits diverse and complex features, encompassing various sound spectra and speech characteristics, which blur the distinction between laughter and other emotions, increasing the difficulty of laughter recognition (Trigeorgis et al., 2016). Additionally, as a form of emotional expression, laughter recognition often necessitates considering more contextual information and comprehensive factors, further enhancing the complexity and challenges of laughter recognition.

## 2. Research Questions

Given the importance of laughter in emotional communication, social connections, and mental health, as well as the challenges faced by current laughter recognition methods, including insufficient datasets, complex and diverse features, and poor comprehensive emotional recognition, this paper aims to explore a more efficient and accurate laughter recognition method to improve its accuracy and practicality. In response to these issues, the following solutions are proposed: firstly, integrating multi-feature extraction techniques. Combining spectral features, temporal features, and other acoustic feature extraction techniques (Nesakumari, 2022), the complex characteristics of laughter are comprehensively captured, enhancing the accuracy and comprehensiveness of feature extraction. Secondly, employing deep learning techniques. Efficient classifier models are constructed to achieve accurate recognition and classification of laughter by using deep learning techniques (Gosztoly et al., 2016).

## 3. Laughter Recognition System Design

### 3.1. System interface

This study aims to achieve laughter recognition by integrating image processing and audio processing technologies. The system provides functions for loading video files, extracting image features, extracting audio features, detecting facial expressions, and extracting emotional features from laughter to achieve accurate laughter recognition and emotional analysis. We have designed and implemented a laughter recognition system, as shown in Figure 1.

*Figure 1*. System Interface

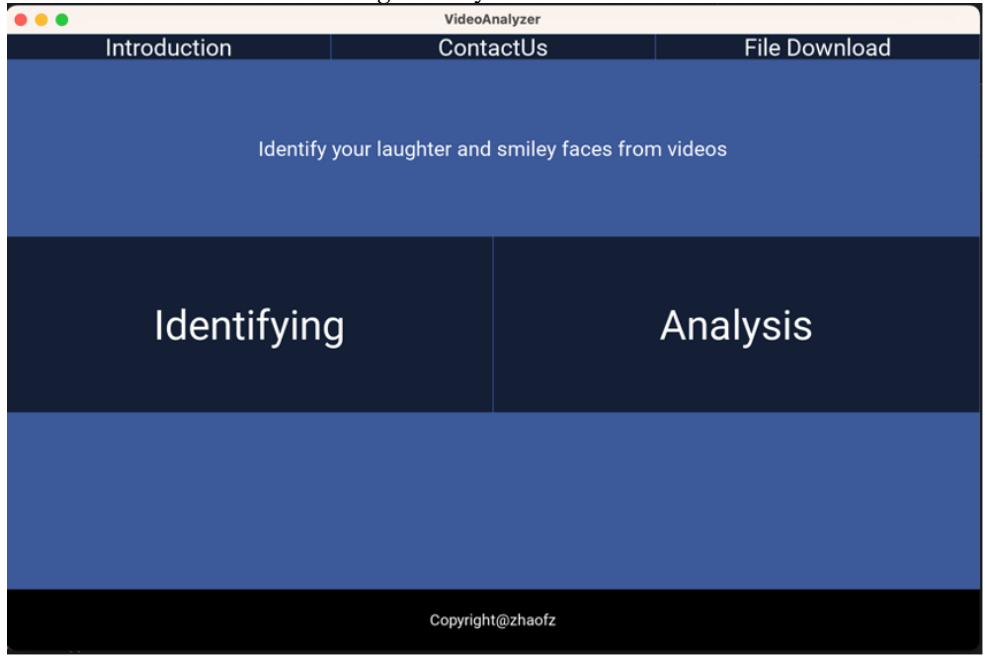

### 3.2. System development challenges and solutions

In the process of developing this system, a complex and multi-functional system, many technical challenges were inevitably encountered. These challenges are not only related to the underlying data processing and feature extraction, but also cover model training and optimization, user interaction design and other aspects. In order to build an efficient, accurate, and user-friendly system, the study analyzes these challenges in depth and develops corresponding solution strategies, as shown in Table 1.

*Table 1*. System development challenges and solutions

| Challenges | Solutions |
| --- | --- |
| Cross-modal Feature Fusion | Feature Extraction Algorithms, Multi-modal Fusion, CNN Integration |
| Video Processing Techniques | OpenCV Library, Parallel Processing, Optimized Algorithms, Grayscale |
|  | Conversion, Frame Loading |
| Audio Processing Techniques | librosa Library, MFCC Feature Extraction, Frequency Range, Energy |
| Model Generalization | Distribution, Temporal Characteristics |
|  | Data Augmentation, Regularization, Transfer Learning, Adaptability Improvement |
| User Interface Design & Experience Optimization | User Research, Interface Design Principles, Simplicity & Intuitiveness, Ease of Use, User Experience, Satisfaction Improvement |

As shown in the table, the first challenge is Cross-modal Feature Fusion. The recognition of laughter does not only rely on a single piece of audio or video information, but needs to integrate the cross-modal features between audio and video (Tanaka & Campbell, 2014). However, how to effectively fuse these features from different modalities is an important challenge in system development. To this end, the research designs specialized feature extraction algorithms and utilizes multimodal fusion deep learning models, such as those incorporating convolutional neural networks (CNNs) (Jogin et al., 2018), to achieve the fusion of audio and video features. This approach can fully utilize the complementary information between different modalities to improve the accuracy of laughter recognition.

The second challenge is Video Processing Techniques. Video processing is a key aspect of the system, which includes multiple steps such as frame loading, grayscale conversion, and face detection. These steps are computationally intensive and require high real-time performance, which brings challenges to the system performance (Sharma et al., 2021). For this reason, we adopt OpenCV, an efficient image-processing library (Culjak et al., 2012), to improve the efficiency of video processing through parallel processing and optimization algorithms.

The third challenge is Audio Processing Techniques. Audio processing also faces many challenges, such as how to accurately extract the MFCC (Mel Frequency Cepstrum Coefficients) features of laughter (Truong & Van Leeuwen, 2007), and how to extract features based on the frequency range, energy distribution, and time-domain characteristics of laughter (Szameitat et al., 2011). To this end, librosa, a specialized audio processing library, is used to extract the MFCC features of the audio (Eyben, Wöllmer & Schuller, 2010). At the same time, we designed a specialized feature extraction algorithm based on the acoustic properties of laughter to ensure that the key information about laughter can be captured.

The fourth challenge is Model Generalization. The ability of a model to generalize under different environments and conditions is an important indicator of system performance. However, due to the complexity and diversity of real-world environments, models are prone to overfitting or underfitting problems (Miüller et al., 2020). For this reason, research has been conducted to improve the model's generalization ability through data augmentation, regularization, and transfer learning. Data augmentation increases the amount of training data for the model and improves its adaptability to different data (Rebuffi et al., 2021); regularization prevents the model from overfitting (Ying, 2019); and migration learning accelerates the training process and improves the performance of the model by using the pre-trained model.

The fifth challenge is User Interaction Design, which is directly related to user experience and satisfaction. How to design a simple, intuitive, and user-friendly interface is an important part of system development (Xu et al, 2021). To this end, we use user research and interface design principles to guide user interaction design. The research deeply understands the users' needs and habits, and combined with the interface design principles, we design a simple, intuitive and easy-to-operate interface. At the same time, we focus on the collection and analysis of user feedback in order to continuously optimize and improve the interface design.

### 3.2. System functions

The system comprises several key modules that extract facial information from videos and laughter features from the audio, ultimately achieving accurate laughter recognition and classification. The functionalities of the system modules are shown in Figure 2. As shown in the figure, the system is a comprehensive system specialized in processing and analyzing laughter, and its core objective is to achieve efficient and accurate laughter recognition by deep processing of video and audio data. The system mainly includes the following key modules and processes.

Data preparation module. This module is the starting point of the system and is responsible for obtaining raw data from the MAHNOB Laughter database (Petridis, Martinez & Pantic, 2013). These data may include video files and audio files, which are initially processed and formatted by the system to ensure that they are suitable for use by the subsequent video processing and audio processing modules.

*Figure 2.* System Founctions

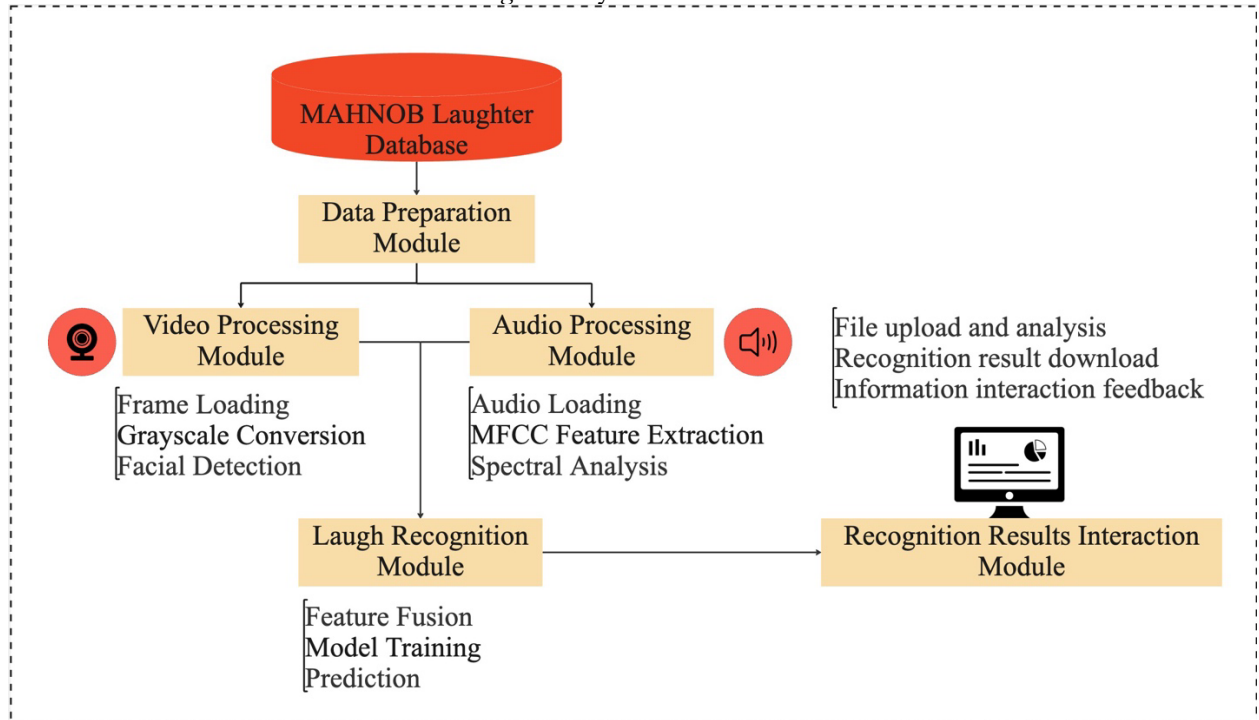

Video Processing Module. This module is responsible for loading video files and performing the necessary video frame loading and conversion. This may include preprocessing steps such as converting video frames to grayscale images for subsequent face detection or feature extraction.

Audio processing module. Parallel to the video processing module, the audio processing module is responsible for loading audio files and performing audio feature extraction. The use of MFCC technique is particularly mentioned here to extract specific features of audio. MFCC is a feature representation method widely used in speech recognition and audio processing, which can effectively capture the frequency and energy distribution information of audio signals.

Feature Extraction Module. This module integrates feature extraction for video and audio. In audio, in addition to MFCC feature extraction, spectral analysis, i.e., spectral analysis, is performed to further resolve the frequency components of the audio signal. On the video side, although the specific video feature extraction method is not explicitly mentioned in the figure, it can be speculated that it may include steps such as face detection to recognize facial expressions in the video, especially the expression features related to laughter.

Feature fusion and model training: after extracting features from video and audio, the system may fuse these features to form a more comprehensive description of laughter. These fused features are then used to train a laughter recognition model. Model training is an iterative process to improve the accuracy and robustness of laughter recognition by continuously optimizing the model parameters.

Recognition Result Interaction Module. The trained model is used for laughter recognition on new video and audio data. The recognition results are uploaded to the server through this module and can be made available to the user for download. In addition, this module is also responsible for receiving feedback from users for further optimization and improvement of the system. In summary, the system realizes efficient recognition of laughter by integrating video processing and audio processing techniques. The system forms a complete processing flow from data preparation to the interaction of recognition results, which provides a strong support for laughter research and application.

## 4. Laughter Recognition System Development Steps

To realize the aforementioned technical solutions, we designed a series of detailed implementation steps covering the entire process from data preparation, image and audio processing, to feature extraction, model design, and training. The specific implementation steps are as below. To streamline the analysis of videos in this study, a systematic approach is followed that seamlessly integrates various steps. Firstly, video files are loaded using the loadVideo method, leveraging the VideoCapture class from the OpenCV library. This enables efficient handling of video data. Subsequently, image features are extracted by iterating through each frame and converting them to grayscale images. This simplifies subsequent tasks such as smile detection and feature extraction. Concurrently, audio features are extracted from the video files through the extractAudioFeatures method. This method utilizes the TarsosDSP library to extract critical audio attributes, including MFCC, which are then stored for comprehensive analysis.

To further enhance audio processing capabilities, the study employs the librosa library for detailed audio analysis, particularly for laughter detection. The librosa.feature.mfcc function is used to compute the MFCC features of the audio, which serve as the foundation for laughter feature extraction. This extraction process focuses on the laughter's distinct frequency range, energy distribution, and temporal characteristics. In parallel, smile detection is performed on the video frames using the detectSmile method with OpenCV's Haar cascade classifier. This classifier accurately identifies smile regions on the grayscale images, facilitating their identification. Then, the study integrates all these methods through the detectLaugh method to recognize laughter in the video files. This comprehensive approach ensures that both visual (smiles) and auditory (laughter) cues are thoroughly analyzed, thereby enhancing the accuracy and reliability of the laughter recognition system. Finally, train the laughter recognition model by the deep learning method.

## 5. System Evaluation and Analysis

This section details the evaluation results of the laughter recognition model. To validate the model's effectiveness and robustness, we employed multiple performance metrics, including accuracy, precision, recall, F1 score, and the confusion matrix. Additionally, we analyzed the model architecture and parameters to ensure the robustness and generalization capability of its performance.

On the test dataset, the model achieved an accuracy of 80%, indicating a high level of correctness in recognizing laughter. The precision was 80%, meaning that 80% of the instances predicted as laughter were indeed true laughter. The recall (sensitivity) was also 80%, indicating that 80% of the actual laughter instances were correctly identified by the model. The F1 score, which is the harmonic mean of precision and recall, was also 80%, demonstrating a balance between accuracy and completeness.

The confusion matrix provides detailed results of the model's performance in classifying laughter and non-laughter instances. Out of 15 test samples, the model correctly identified 12 laughter instances (true positives) and misclassified 3 non-laughter instances as laughter (false positives). The model's performance in predicting the label as 1 (i.e., laughter) was lacking, as reflected by the false negative count of 0, leading to a recall of 0 in the test data.

The model architecture includes multiple convolutional neural network layers and fully connected layers, with a total parameter count of 174,728 and trainable parameters amounting to 58,242. During training, the model's performance progressively improved, with the training accuracy increasing from an initial 60.53% to 69.25%. This indicates that the model adapted to the data over the training process and improved its classification performance.

In summary, the model demonstrated a trend of gradual improvement during training and ultimately achieved an accuracy of 80% on the test dataset. Despite underperformance in certain instances, the model overall exhibited robustness and the ability to handle variations in real-world data. Future work will focus on optimizing feature extraction and model architecture to further enhance the precision and recall of laughter recognition. The high accuracy, precision, recall, and F1 score validate the model's potential application in laughter recognition and emotion analysis, providing users with accurate and reliable analysis results.

## 6. Conclusion

This study designed and implemented a laughter recognition system based on multimodal fusion and deep learning, effectively enhancing the accuracy and robustness of laughter recognition by leveraging image and audio processing techniques. The results indicate that the multimodal fusion approach comprehensively captures laughter features, reducing errors associated with a single modality and thereby improving recognition performance.

Firstly, in data preparation, we successfully connected to the database and extracted a substantial amount of MP4 video and WAV audio data, providing a solid foundation for model training. Subsequently, using the OpenCV library for image processing, we accurately located and extracted smile information from the videos. Additionally, the audio processing module utilized the librosa library, particularly MFCC features, to support laughter emotion analysis. In the laughter recognition module, we fused image and audio features and employed deep learning techniques to achieve accurate laughter classification. Model evaluation results show that the final model achieved 80% accuracy, precision, and recall on the test dataset, with an F1 score of 80%, demonstrating a good balance between accuracy and completeness.

Although the model underperformed in certain instances, it generally exhibited robustness and the ability to handle variations in real-world data. Further optimization of feature extraction algorithms and model architecture is expected to enhance the model's precision and recall. This study not only achieved significant breakthroughs in laughter recognition technology but also laid a foundation for future development and application. Future work will focus on optimizing feature extraction and model architecture to improve recognition accuracy and expand application scenarios. Additionally, this study emphasizes the importance of integrating multimodal fusion methods to enhance the effectiveness of laughter recognition. The laughter recognition technology developed in this study has potential applications in fields such as mental health monitoring, educational activity evaluation, and human-computer interaction, offering new opportunities for emotion analysis and interaction technology development.

Future research directions will focus on optimizing feature extraction algorithms and model architecture, particularly enhancing the performance and accuracy of the recognition model. Specifically, we aim to address the challenges posed by complex environments and diverse laughter forms by leveraging advanced deep learning techniques and multimodal fusion methods. Further exploration will be conducted to identify and analyze the correlation between different features, providing a more comprehensive understanding of laughter recognition. In addition to improving recognition accuracy, future work will expand the application scenarios of laughter recognition technology. Potential applications include mental health monitoring, educational activity evaluation, and human-computer interaction, where laughter recognition can provide valuable insights into emotional states and interaction patterns. Furthermore, we will explore the integration of laughter recognition technology with other affective computing methods to develop more sophisticated emotion analysis systems, enhancing the user experience and promoting the development of emotion recognition technology in various domains.

## References

Provine, R. R. (2001). Laughter: A scientific investigation. Penguin.

Martin, R. A., & Ford, T. (2018). *The psychology of humor: An integrative approach*. Academic press.

Kuiper, N. A., & Martin, R. A. (1998). Laughter and stress in daily life: Relation to positive and negative affect. *Motivation and emotion*, *22*, 133-153.

Goddard, R. D. (2003). The impact of schools on teacher beliefs, influence, and student achievement. *Teacher beliefs and classroom performance: The impact of teacher education*, *6*, 183-202.

Martin, R. A., Puhlik-Doris, P., Larsen, G., Gray, J., & Weir, K. (2003). Individual differences in uses of humor and their relation to psychological well-being: Development of the Humor Styles Questionnaire. *Journal of research in personality*, *37*(1), 48-75.

Eyben, F., Wöllmer, M., & Schuller, B. (2010, October). Opensmile: the munich versatile and fast open-source audio feature extractor. In *Proceedings of the 18th ACM international conference on Multimedia* (pp. 1459-1462).

Turker, B. B., Yemez, Y., Sezgin, T. M., & Erzin, E. (2017). Audio-facial laughter detection in naturalistic dyadic conversations. *IEEE Transactions on Affective Computing*, *8*(4), 534-545.

Trigeorgis, G., Ringeval, F., Brueckner, R., Marchi, E., Nicolaou, M. A., Schuller, B., & Zafeiriou, S. (2016, March). Adieu features? end-to-end speech emotion recognition using a deep convolutional recurrent network. In *2016 IEEE international conference on acoustics, speech and signal processing (ICASSP)* (pp. 5200-5204). IEEE.

Tanaka, H., & Campbell, N. (2014). Classification of social laughter in natural conversational speech. *Computer Speech & Language*, *28*(1), 314-325.

Sharma, V., Gupta, M., Kumar, A., & Mishra, D. (2021). Video processing using deep learning techniques: A systematic literature review. *IEEE Access*, *9*, 139489-139507.

Miüller, J. C., Weibel, R., Lagrange, J. P., & Salgé, F. (2020). Generalization: state of the art and issues. *GIS And Generalisation*, 3-17.

Xu, W., Dainoff, M. J., Ge, L., & Gao, Z. (2021). From human-computer interaction to human-AI Interaction: new challenges and opportunities for enabling human-centered AI. *arXiv preprint arXiv:2105.05424*, *5*.

Nesakumari, G. R. (2022). Image retrieval system based on multi feature extraction and its performance assessment. *International Journal of Information Technology*, *14*(2), 1161-1173.

Gosztolya, G., András, B. E. K. E., Neuberger, T., & László, T. Ó. T. H. (2016). Laughter classification using Deep Rectifier Neural Networks with a minimal feature subset. *Archives of Acoustics*, *41*(4), 669-682.

Jogin, M., Madhulika, M. S., Divya, G. D., Meghana, R. K., & Apoorva, S. (2018, May). Feature extraction using convolution neural networks (CNN) and deep learning. In *2018 3rd IEEE international conference on recent trends in electronics, information & communication technology (RTEICT)* (pp. 2319-2323). IEEE.

Culjak, I., Abram, D., Pribanic, T., Dzapo, H., & Cifrek, M. (2012, May). A brief introduction to OpenCV. In *2012 proceedings of the 35th international convention MIPRO* (pp. 1725-1730). IEEE.

Szameitat, D. P., Darwin, C. J., Szameitat, A. J., Wildgruber, D., & Alter, K. (2011). Formant characteristics of human laughter. *Journal of voice*, *25*(1), 32-37.

Truong, K. P., & Van Leeuwen, D. A. (2007). Automatic discrimination between laughter and speech. *Speech Communication*, *49*(2), 144-158.

Rebuffi, S. A., Gowal, S., Calian, D. A., Stimberg, F., Wiles, O., & Mann, T. A. (2021). Data augmentation can improve robustness. *Advances in Neural Information Processing Systems*, *34*, 29935-29948.

Ying, X. (2019, February). An overview of overfitting and its solutions. In *Journal of physics: Conference series* (Vol. 1168, p. 022022). IOP Publishing.

Petridis, S., Martinez, B., & Pantic, M. (2013). The MAHNOB laughter database. *Image and Vision Computing*, *31*(2), 186-202.